\documentclass[12pt]{article}
\usepackage{scicite}
\usepackage{times}
\usepackage{graphicx}

\newcommand{\ga}{\ifmmode\stackrel{>}{_{\sim}}\else$\stackrel{>}{_{\sim}}$\fi}
\newcommand{\la}{\ifmmode\stackrel{<}{_{\sim}}\else$\stackrel{<}{_{\sim}}$\fi}

\newcommand{\msun}{\ifmmode\mbox{M}_{\odot}\else$\mbox{M}_{\odot}$\fi}
\newcommand{\lsun}{\ifmmode\mbox{L}_{\odot}\else$\mbox{L}_{\odot}$\fi}
\newcommand{\rsun}{\ifmmode\mbox{R}_{\odot}\else$\mbox{R}_{\odot}$\fi}

\newcommand{\terad}{PSR~J1748$-$2446ad}
\newcommand{\terA}{PSR~J1748$-$2446A}
\newcommand{\terO}{PSR~J1748$-$2446O}

\newcommand{\ngcpsr}{PSR~J1740$-$5340}

\newenvironment{sciabstract}{%
\begin{quote} \bf}
{\end{quote}}

\newcounter{lastnote}
\newenvironment{scilastnote}{%
\setcounter{lastnote}{\value{enumiv}}%
\addtocounter{lastnote}{+1}%
\begin{list}%
{\arabic{lastnote}.}
{\setlength{\leftmargin}{.22in}}
{\setlength{\labelsep}{.5em}}}
{\end{list}}

\title{A Radio Pulsar Spinning at 716\,Hz}

\author
{Jason W.~T.~Hessels$^{1\ast}$, Scott M.~Ransom$^{2}$, Ingrid
H.~Stairs$^{3}$,\\ Paulo C.~C.~Freire$^{4}$, Victoria M.~Kaspi$^{1}$, and Fernando Camilo$^{5}$ 
\\
\normalsize{$^{1}$Department of Physics, McGill University, Montreal, QC H3A 2T8, Canada}\\
\normalsize{$^{2}$NRAO, 520 Edgemont Rd., Charlottesville, VA 22903, USA}\\
\normalsize{$^{3}$Department of Physics and Astronomy, University of British Columbia,} \\
\normalsize{6224 Agricultural Road, Vancouver, BC V6T 1Z1, Canada}\\
\normalsize{$^{4}$NAIC, Arecibo Observatory, HC03 Box 53995, PR 00612, USA}\\
\normalsize{$^{5}$Columbia Astrophysics Laboratory, Columbia University, 550 West 120th Street,}\\
\normalsize{New York, NY 10027, USA}\\
\normalsize{$^\ast$To whom correspondence should be addressed; E-mail: hessels@physics.mcgill.ca}
}

\date{}

\begin{document} 
\baselineskip24pt
\maketitle 

\begin{sciabstract}
  We have discovered a 716-Hz eclipsing binary radio
  pulsar in the globular cluster Terzan~5 using the Green Bank
  Telescope.  It is the fastest-spinning neutron star ever found,
  breaking the 23-year-old record held by the 642-Hz pulsar
  B1937+21. The difficulty in detecting this pulsar, due
  to its very low flux density and high eclipse fraction ($\sim$40\%
  of the orbit), suggests that even faster-spinning neutron stars
  exist.  If the pulsar has a mass less than 2\,\msun, then its
  radius is constrained by the spin rate to be $< 16$\,km. The short
  period of this pulsar also constrains models that suggest
  gravitational radiation, through an r-mode instability,
  limits the maximum spin frequency of neutron stars.
\end{sciabstract}

While the majority of neutron stars are observed to rotate slower than
a few times a second, those in binary systems can reach spin rates of
hundreds of times a second through the transfer of angular momentum
from their companion star\cite{acrs82,specialrs82}.  Some of these
neutron stars, termed millisecond pulsars, are persistent radio
sources whose emission is modulated at the star's spin frequency.
Determining the maximum achievable rotation rate of a neutron star is
important for a variety of astrophysical problems, ranging from
understanding the behaviour of matter at supra-nuclear densities, to
estimating the importance of neutron stars as gravitational wave
sources for current and upcoming gravitational wave detectors.  For
over 23 years, the 642-Hz pulsar B1937+21, the first millisecond
pulsar ever found, has been the fastest-spinning neutron star
known\cite{bkh+82}.  It has been argued that faster ones are
exceedingly rare, if they exist at all\cite{cmm+03}.

Per unit mass, globular clusters have many more millisecond pulsars than the
Galactic disk.  This is due to the extremely high stellar densities in their
cores ($10^4 - 10^6$\,pc$^{-3}$), which promote the creation of binary
systems\cite{mh97} where a neutron star is spun-up (or ``recycled'') to
rotate hundreds of times a second\cite{acrs82,specialrs82}.  We have
searched the massive and dense globular cluster Terzan~5 for millisecond
pulsars using the National Radio Astronomy Observatory's \cite{nrao} 100-m
Green Bank Telescope (GBT). Our searches have thus far uncovered 30
millisecond pulsars in Terzan~5\cite{pauloweb}, in addition to the three
previously known pulsars in this cluster\cite{ljm90,lmbm00,ran01}. The
discovery of 21 pulsars in Terzan~5 was presented in \cite{rhs+05}.
Following that paper, an additional nine pulsars have been found in searches
of our monitoring observations. These will be reported elsewhere.  Terzan~5
has the largest known population of millisecond pulsars of any globular
cluster, roughly one quarter of the entire population of globular cluster
millisecond pulsars, and the five fastest rotating pulsars in the Galactic
globular cluster system\cite{pauloweb}.  Among the newest discoveries
in Terzan~5 is \terad, a 716-Hz eclipsing binary millisecond pulsar, which
is the fastest-spinning neutron star ever found.


{\bf Observations and data analysis.} We discovered \terad\ in 10 November
2004 observations of Terzan~5 and confirmed it in 8 January 2005
observations using the Pulsar Spigot backend \cite{kel+05} on the GBT.  All
observations employed the Spigot with 600 MHz of usable bandwidth centered at
1950\,MHz, 768 spectral channels, and 81.92-$\mu$s
sampling.  Observations were generally 6$-$7 hours in length and taken at
roughly monthly intervals starting June 2004.  In addition, there was a more
closely spaced set of observations in early May 2005.

The discovery observation showed that the pulsar is part of a binary system
and is eclipsed by its companion; both of these properties restricted our
ability to detect the pulsar in our monitoring sessions.  Nonetheless, we
have now detected the pulsar in at least 18 out of the 30 multi-hour
observations taken thus far (see Fig 1 for pulse profile). We have derived a
reliable orbital ephemeris (see Tab~1) by initially modelling the pulse
phase delays of a few good detections with a simple sine function and then
refining the model by fitting pulse times of arrival to a simple Keplerian
orbital model, with arbitrary pulse phase offsets between observing epochs.
This ephemeris allowed us to detect the pulsar in many observations where it
was not initially identified through a periodicity search.

The pulsar is in a highly circular 26-hr orbit with a \ga0.14\,\msun\
companion, and is eclipsed for $\sim$40\% of its orbit at 2\,GHz. 
Such a large eclipse fraction, corresponding to an eclipse region with
physical size $\sim$5$-$6\,\rsun\ is extremely rare for such a relatively
wide orbit (separation between the pulsar and companion of
$\sim$4$-$5\,\rsun).  The companion may be a bloated main-sequence star,
possibly still filling its Roche Lobe, as has been suggested for \ngcpsr, a
35-hr binary millisecond pulsar with a \ga0.21\,\msun\ companion and
$\sim$40\% eclipse fraction at 1.4\,GHz\cite{dpm+01}.  The eclipse
properties are also similar to those of \terA, a 1.8-hr binary with a
0.089\,\msun\ minimum mass companion, also located in Terzan~5\cite{ljm90}.
Like \ngcpsr\ and \terA, there is evidence that the eclipse duration of
\terad\ is variable, and sometimes lasts longer than 40\% of the orbit.  On
a least two epochs when our ephemeris predicts the pulsar should have been
visible for several hours, it was not detected at all.  The pulse
signal-to-noise is too low to measure dispersion measure variations on short
timescales.  Future observations should allow a phase-coherent timing
solution to be derived, which will provide a precise position and observed
spin frequency derivative $\dot{\nu}$.  Until then, we have provided an
upper limit on $|\dot{\nu}|$ (Tab~1).

We have verified that this signal is not harmonically related to any
of the other known pulsars in Terzan~5.  We have also carefully
investigated the possibility that our searches have identified the
second harmonic of a new 358-Hz pulsar.  When we fold the data at
358\,Hz, there are two identical peaks (within the resolution and RMS
noise level of the Spigot data) separated by 180$^{\circ}$ in pulse
phase.  This is what we expect to see if the pulsar signal is folded
at half its intrinsic spin frequency, and strongly suggests that
716\,Hz is the true spin frequency of the pulsar.  The results of
folding the data at other harmonically related spin frequencies are
also consistent with the pulsar having a true frequency of 716\,Hz.
There is also evidence (Fig 1) for a weak, but statistically
significant interpulse (extra structure in addition to the main pulse
peak) when the data are folded at 716\,Hz. This interpulse, if real, is 
further evidence that the spin frequency is 716\,Hz.

Lastly, we have simultaneously observed the pulsar using the GASP
coherent dedispersion pulsar machine \cite{drb+04,fsb+04}, which
records only $\sim$$1/6$ the bandwidth achievable with Spigot but
which removes all dispersive smearing due to the ionized interstellar
medium, resulting in significantly sharper (i.e. narrower) pulse
profiles\cite{spigot}.  When the GASP data are folded at 358\,Hz, two
peaks, consistent in shape with each other to within the RMS of the
noise, are seen separated by 180$^{\circ}$ in pulse phase, again
indicating that 358\,Hz is half the true spin frequency. We conclude
that \terad\ is indeed a 716-Hz pulsar.


{\bf Implications and discussion.} The equation of state of matter at
supra-nuclear densities, and thus the mass-radius relation for neutron
stars, is unknown.  If a star is rotating too rapidly for a given radius,
centrifugal forces will cause it to shed mass.  Lattimer \& Prakash
\cite{lp04} derive the following equation which, independent of the true
equation of state, gives the maximum spin frequency for a neutron star with
a non-rotating radius $R$ and mass $M$ (assuming this is not close to the
maximum mass allowed by the equation of state): $\nu_{max} = 1045
(M/\msun)^{1/2}(10\rm{km}/R)^{3/2}$\,Hz.  Using this, and assuming a mass
less than 2\,\msun\ (which accomodates all measured neutron star
masses\cite{lp04}) we find an upper limit of 16\,km on \terad's radius.  We
note that this constraint applies specifically to \terad\ and that
slower-spinning pulsars could have larger radii.  Recently, Li \& Steiner
\cite{ls05} have derived a radius range of 11.5$-$13.6\,km for a 1.4\,\msun\
neutron star, based on terrestrial laboratory measurements of nuclear
matter.  For a 1.4\,\msun\ neutron star, we find an upper limit of 14.4\,km,
which is in agreement with their result.  These radius constraints are more
robust than those obtained through observations of neutron star thermal
emission, which is faint, difficult to measure, and whose characterization
depends on uncertain atmosphere models\cite{krh05}.  Although in principle
a radius measurement could constrain the unknown equation of state of dense
matter, \terad\ does not rule out any particular existing models, since the
pulsar mass is unknown.  It is unlikely that a mass measurement will be
achievable through timing of the pulsar, as the orbit is too circular to
measure the relativistic advance of the periastron, which would likely be
contaminated by classical effects as well.  However, once a timing position is
available, optical spectroscopy may allow determinarion 
of the mass ratio, as has been done in the case of \ngcpsr\ in the
globular cluster NGC~6397\cite{fsg+03}.  The feasibility of such a
measurement will depend on the pulsar not being located in the dense and
optically crowded core of the cluster.  Fortunately, as \terad\ has a
dispersion measure which is $\sim 3$ units lower than the $\sim 239$\,pc
cm$^{-3}$ average dispersion measure of the cluster pulsars, it is likely to
be located outside of the core.

Although there are selection effects, especially at radio wavelengths,
against finding fast, binary pulsars \cite{jk91}, we have maintained
excellent sensitivity to these through the use of advanced search techniques
\cite{rem02} and data with high time and frequency resolution.  For example,
these searches blindly detect the 596-Hz binary pulsar \terO\ ($P_{orb} \sim
6$\,hrs) at its second and fourth harmonic, which is equivalent to detecting
a highly accelerating 1192-Hz pulsar and its second harmonic.  While these
searches are clearly sensitive to pulsars faster than 716\,Hz, we note that
obscuration of the pulsar signal by material blown off the companion by the
pulsar wind may play an important role in reducing the chances of detecting
such systems\cite{tav91}.  Of the five fastest known millisecond pulsars
(including the two found in the Galactic plane; see Tab~2), four are
eclipsing, and the fifth, PSR B1937+21, is unusual because, unlike
$\sim$80\% of Galactic plane millisecond pulsars, it has no binary
companion.  Since rotational energy loss is inversely proportional to the
cube of the spin period ($\dot{E} = 4 {\pi}^2 I \dot{P}/P^3$, where
$\dot{E}$ is the spin-down luminosity and $\dot{P}$ is the period
derivative), it is plausible that a large fraction of the fastest binary
pulsars are evading detection because their powerful winds are ablating
their companions.  Some of the ablated material remains in the vicinity of
the system and obscures radio pulsations, particularly at lower radio
frequencies.  Although we have no detections of the pulsar at other
frequencies, several other eclipsing pulsars are observed to have longer
eclipse durations at lower radio frequencies\cite{rsb+04}.  If this is also
the case for \terad, then the unusually high observing frequency used here
(2\,GHz, while most globular cluster surveys have been conducted at 1.4\,GHz
or lower) was likely crucial in detecting this pulsar.  \terad\ is too weak
(the flux density at 1950\,MHz is $\sim$0.08\,mJy) to be detectable by the
vast majority of Galactic plane surveys and its high eclipse fraction
compounds this problem. This suggests that other even faster-spinning
pulsars exist, but will require deeper surveys (perhaps at higher observing
frequencies to mitigate obscuration of the pulsar signal by intra-binary
material) and more concentrated efforts to be detected. In effect, the
isolated nature and very large luminosity of PSR B1937+21 make it a unique
object, rather than a representative member of the millisecond pulsar
population.

Low-mass X-ray binaries (LMXBs) with neutron star members are the
likely progenitors of the radio millisecond pulsars.  As the spin-up
time for a neutron star to reach $> 1000$\,Hz rotation via the
accretion of matter in an LMXB is much shorter than the typical LMXB
lifetime, one might naively expect many millisecond pulsars to be rotating
at sub-millisecond periods. However, given the observed lack of
pulsars spinning this fast, gravitational radiation has been proposed
as a limiting mechanism, as it could be responsible for carrying away
rotational kinetic energy from the star, thus spinning it down.
Specifically, gravitational waves may be acting either through an
accretion-induced mass quadrupole on the crust \cite{bil98}, a large
toroidal magnetic field \cite{cut02}, or an r-mode (Rossby wave)
instability in the stellar core \cite{aks99,lu01}.

\terad\ provides interesting constraints on the r-mode possibility.
These oscillations are believed to be present in all rotating neutron
stars \cite{and98,fm98}.  Due to gravitational radiation emission, the
r-modes become unstable and grow exponentially.  The amplitude of the
mode continues to grow as angular momentum is radiated away, and the
star spins down.  However, it is unclear whether the driving of these
modes by gravitational waves can overcome viscous damping in the star.
Damping depends on the core temperature of the star and its spin rate,
as well as several other factors including the thickness of the
neutron star crust and how it couples to the core. For a given core
temperature, it is possible to derive a critical spin frequency above
which the proposed mode is unstable and will cause the star to spin
down rapidly.  It has been predicted that the critical frequency is $<
700$\,Hz for a wide range of core temperature ($10^7 - 10^{10}$\,K)
and a realistic model of the neutron star crust \cite{lu01}.  Our
discovery of a 716-Hz pulsar indicates that if the r-mode instability
limits neutron star spin-up, then either it must become important only
at more rapid spin rates or better modelling of the neutron star crust
is required.  The current upper limit on the frequency derivative of
\terad\ is consistent with those measured for other millisecond
pulsars, and does not suggest an anomalously rapid spin-down rate.
Thus, while there is currently no evidence that \terad\ is spinning
down due to gravitational radiation, the possible importance of
systems like \terad\ as gravitational wave sources for detectors like
LIGO makes improved modelling of neutron star structure and gravitational
wave instabilities critically important.

The observed spin frequencies of LMXBs are all less than 620\,Hz. The biases
that exist at radio wavelengths against finding much faster pulsars do not
exist for LMXBs, as X-rays do not suffer from the dispersive effects of the
interstellar medium and are less obscured by intra-binary material.  This
suggests that faster-spinning neutron stars in LMXBs should be detectable if
they exist.  However, transient coherent pulsations are only observed in
seven sources, and there may be unidentified sources of bias in the
population that are preventing faster pulsations from being detected.
Chakrabarty et al. 2003 \cite{cmm+03} performed a Bayesian statistical
analysis on the spin frequencies of 11 nuclear-powered millisecond pulsars,
those for which the spin frequency is known from the detection of burst
oscillations, and concluded that the sample is consistent (at the 95\%
confidence level) with a cutoff $\nu_{max} = 760$\,Hz.  More recent work by
the same authors \cite{cha04}, which added 2 pulsars to the sample, has
revised this limit to 730\,Hz. Based on this, they conclude that something,
possibly gravitational radiation, is limiting the spin frequency of neutron
stars. If their statistically derived upper limit is realistic (and
therefore some mechanism is limiting neutron star spin-up), then the 716-Hz
pulsar presented here is likely an extremely rare object. However, we note
that their Bayesian calculation is very sensitive to the neutron star with
the slowest spin frequency included in the analysis.  In addition, the
assumption made by these authors that the pulsar spin rates are uniform in
frequency (at least over some range $\nu_{low}$ to $\nu_{high}$) would
likely not apply to the Terzan~5 pulsars, whose spin period distribution is
clearly not uniform\cite{rhs+05}, even accounting for the observational bias
against detecting the fastest pulsars.  Hence, a recalculation of the
maximum spin cutoff using the Terzan~5 sample of pulsars and the same
statistical analysis would not be appropriate, although the existence of
\terad\ already implies that the cutoff must be higher.

\begin{scilastnote}
\item We thank Robert Ferdman and Paul Demorest for help with GASP
  observations, and acknowledge their work, along with Donald Backer,
  David Nice, R.~Ramachandran, and Joeri van Leeuwen in creating the
  GASP instrument. We also thank Lars Bildsten, Phil Arras and Jim
  Lattimer for very useful discussions. JWTH is an NSERC PGS-D fellow
  and is grateful to the Canada Foundation for Innovation for funding
  the computing resources used to make this discovery and to Paul
  Mercure for helping to maintain them.  IHS holds an NSERC UFA and is
  supported by a Discovery grant and UBC start-up funds. VMK is a
  Canada Research Chair and is supported by an NSERC Discovery Grant
  and Steacie Fellowship Supplement, by the FQRNT and CIAR.  FC thanks
  the US NSF for support.  GASP is funded by an NSERC RTI-1 grant to
  IHS and by US NSF grants to Donald Backer and David Nice.
\end{scilastnote}

\clearpage

\begin{table}
\begin{center}
\begin{tabular}{ll}
\hline \hline
Parameter & Value \\
\hline \hline
{\it Rotational Parameters} & \\
\hline
Pulse period $P$ (s)                          & 0.00139595482(6) \\
Period derivative $|\dot{P}|$ (s/s)           & $\le 6 \times 10^{-19}$ \\
Pulse frequency $\nu$ (Hz)                    & 716.35556(3) \\
Frequency derivative $|\dot{\nu}|$ (Hz/s)     & $\le 3 \times 10^{-13}$ \\
Epoch (MJD)                                   & 53500 \\
\hline
{\it Orbital Parameters} & \\
\hline
Orbital period $P_{orb}$ (days)               & 1.09443034(6) \\
Projected semi-major axis $x$ (lt-s)          & 1.10280(6) \\
Time of ascending node $T_{ASC}$ (MJD)        & 53318.995689(12) \\
Eccentricity $e$                              & $< 0.0001$  \\
\hline
{\it Derived Quantities} & \\
\hline
Companion minimum mass $M_{2,min}$ (\msun)    & 0.14 \\
Dispersion measure DM (pc cm$^{-3}$)          & 235.6(1) \\
Flux density at 1950 MHz $S_{1950}$ (mJy)     & 0.08(2) \\
Characteristic age $\tau_{c}$ (years)         & $\ge 2.5 \times 10^7$ \\
Surface magnetic field $B_{surf}$ (G)         & $\le 1.1 \times 10^9$ \\
Spin-down luminosity $\dot{E}$ erg/s          & $\le 1.3 \times 10^{37}$ \\
\hline \hline
\end{tabular}
\caption{Measured and derived parameters of \terad.  All measured spin
  and orbital parameters were determined using the {\tt TEMPO}
  software package \cite{tempo}, using arbitrary phase offsets between
  observing epochs.  Given the currently sparsely sampled data, it is
  impossible to phase connect separate observations.  For this reason,
  we provide only an upper limit on the magnitude of the spin
  frequency derivative of the pulsar, which incorporates the maximum
  possible acceleration due to the gravitational potential of Terzan~5
  assuming a position close to the cluster center.  Likewise, we can
  currently only place limits on the derived characteristic age,
  surface magnetic field and spin-down luminosity.  The minimum
  companion mass is derived assuming a pulsar mass of 1.4\,\msun.
  The dispersion measure was determined by measuring pulse arrival time
  delays in the different frequency channels across the 600-MHz observing
  bandwidth.}
\end{center}
\end{table}

\clearpage

\begin{table*}
\begin{center}
\begin{tabular}{ l c r l c c }
\hline \hline
\multicolumn{1}{c}{} &
\multicolumn{1}{c}{Spin Frequency} &
\multicolumn{1}{c}{$P_b$} &
\multicolumn{1}{c}{$M_{2,min}$} &
\multicolumn{1}{c}{Eclipse} &
\multicolumn{1}{c}{} \\
\multicolumn{1}{c}{Pulsar} &
\multicolumn{1}{c}{(Hz)} &
\multicolumn{1}{c}{(days)} &
\multicolumn{1}{c}{(M$_{\odot}$)} &
\multicolumn{1}{c}{Fraction} &
\multicolumn{1}{c}{Location} \\
\hline \hline
J1748$-$2446ad & 716.358 & 1.0944   &  0.14  & 0.4  & Terzan~5 \\
B1937$+$21     & 641.931 & \multicolumn{3}{c}{isolated} & Galaxy   \\
B1957$+$20     & 622.123 & 0.3819   &  0.021 & 0.1  & Galaxy   \\
J1748$-$2446O  & 596.435 & 0.2595   &  0.035 & 0.05 & Terzan~5 \\
J1748$-$2446P  & 578.496 & 0.3626   &  0.37  & 0.4  & Terzan~5 \\
J1843$-$1113   & 541.812 & \multicolumn{3}{c}{isolated} & Galaxy   \\
J0034$-$0534   & 532.714 & 1.5892   &  0.14  & 0    & Galaxy   \\
J1748$-$2446Y  & 488.243 & 1.17     &  0.14  & 0    & Terzan~5 \\
J1748$-$2446V  & 482.507 & 0.5036   &  0.12  & 0    & Terzan~5 \\
B0021$-$72J    & 476.048 & 0.1206   &  0.020 & 0.1$^*$  & 47~Tucanae \\
\hline \hline
\end{tabular}
\caption{\label{tab:fast_pulsars} The ten fastest-spinning known radio pulsars.
  Data compiled from the ATNF pulsar database\cite{atnfweb}.
  $^*$B0021$-$72J is eclipsed only at radio frequencies $<$1\,GHz.}
\end{center}
\end{table*}

\clearpage

\begin{figure}[t]
\begin{center}
\includegraphics[height=4in,angle=270]{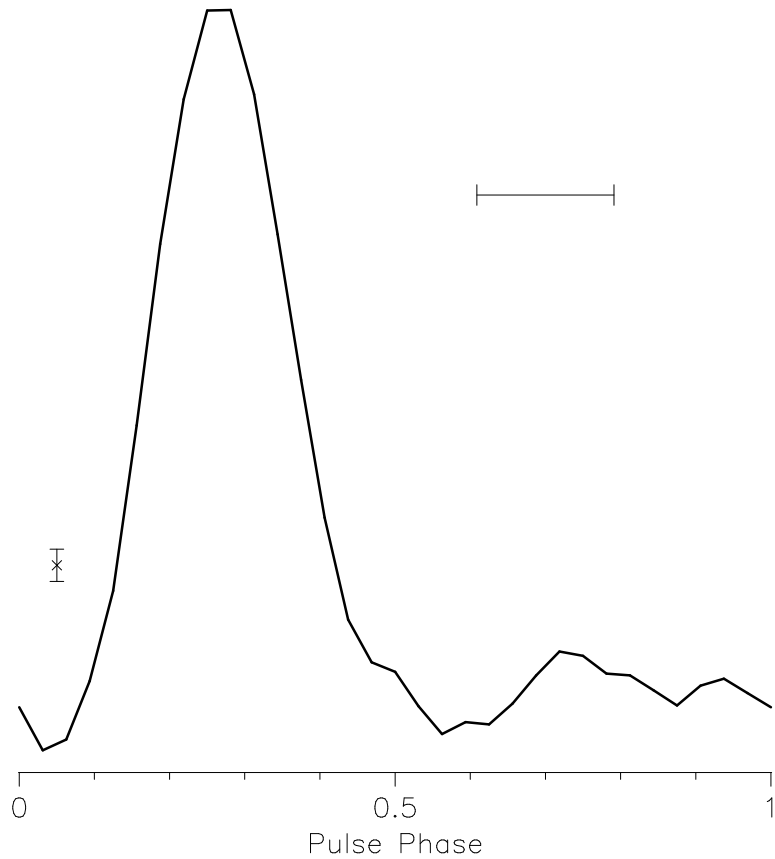}
\end{center}
\caption{Master \terad\ pulse profile from the combination of eight GBT
  Pulsar Spigot observations at 2\,GHz with particularly good detections of
  the pulsar.  The cumulative integration time is $\sim$54\,hrs.  There are
  32 phase bins across the profile, and the y-axis plots flux in arbitrary
  units.  The one sigma error bar on the flux is shown in the lower
  left-hand corner.  The effective time resolution of the data, $\sim
  300$\,$\mu$s, which accounts for pulse smearing due to channelisation of
  the dispersed data and finite time sampling, is indicated by the
  horizontal bar.  A weak, but statistically significant interpulse is seen
  at a phase of $\sim$0.75.}
\end{figure}


\begin{thebibliography}{10}
\bibitem{spigot}
The SPIGOT pulse profile is significantly widened compared with the intrinsic
  pulse width of the pulsar because the finite number of channels used across
  the observing bandwidth does not allow for a complete removal of the
  dispersive effects of the interstellar medium.
\end{thebibliography}
\end{document}